\documentstyle[aps,preprint]{revtex}

\input epsf

\begin{document}
{\tighten 
\title{\bf Zero Modes for the $D\!=\!11$ Membrane and Five-brane}
\author{David M. Kaplan and Jeremy Michelson}
\address{Department of Physics \\ University of California \\
	Santa Barbara, California  93106}
\bigskip
\date{October 9, 1995}
\maketitle
\begin{abstract}
There exist extremal p-brane solutions of $D\!=\!11$ supergravity for
p=2~and~5.  In this paper we investigate the zero modes of the
membrane
and the five-brane solutions as a first step toward understanding
the
full quantum theory of these objects.  It is found that both
solutions possess the correct number of normalizable zero modes
dictated by supersymmetry.
\end{abstract}

\pacs{04.65.+e}
} 

\section{Introduction} \label{intro}
Recent discussions of dualities in string theory indicate that
$D=11$ supergravity may be relevant to string theory.  For example,
it has been argued~\cite{townsend,hull,witten}
that $D=11$ supergravity is the
strong coupling limit of $D=10$ Type IIA string theory.  Coupled
with the conjecture~\cite{townsend,hull,witten} that $D=10$ Type
IIA strings are compactified $D=11$ membranes~\cite{oldduff},
it follows that
$D=11$ supergravity is relevant to
$D=10$ Type IIA theory~\cite{hull,duffetal}.
The study of the $D=11$ membrane and five-brane solutions are
therefore of interest.

In this paper, the effective actions for the membrane and
five-brane solutions of $D=11$ supergravity are calculated from the
$D=11$ supergravity Lagrangian.  In Section~\ref{prelim} we review
the solutions.  In Section~\ref{zero} we discuss the zero modes and
find the effective actions.  In particular, we find the effective
action to agree with the action expected from the general theory of
$p$-branes.  We make some observations about the results in
Section~\ref{conclusions}.

\section{Preliminaries -- The solutions} \label{prelim}

First we discuss the nature of the membrane and five-brane 
solutions~\cite{worldbranes,duff&stelle,guven}.
These p-branes solve the $D\!=\!11$
supergravity equations of motion which follow from the 
Lagrangian~\cite{duff&stelle,supergrav} (omitting fermions)
\begin{eqnarray} \label{superl}
{\cal L} & = & \frac{1}{2} \sqrt{-g}R - \frac{1}{24} \sqrt{-g} 
F_{MNPQ}F^{MNPQ}
\nonumber \\
& & +\frac{4}{(12)^4}\varepsilon^{MNOPQRSTUVW}
F_{MNOP}F_{QRST}A_{UVW}.
\end{eqnarray}

The extreme membrane solution to $D\!=\!11$
supergravity can be written
\cite{duff&stelle}
\begin{equation} \label{duffbrane}
\left\{ \begin{array}{lll} 
{{d}}s^2 & = & \Lambda^{-\frac{2}{3}} \eta_{\mu \nu}
{{d}}x^{\mu} {{d}}x^{\nu} + \Lambda^{\frac{1}{3}} {{d}}y^p
{{d}}y^p
\\
A_{\mu \nu \rho} & = & \pm \frac{1}{2} \varepsilon_{\mu \nu \rho}
\Lambda^{-1}.
\end{array} \right.
\end{equation}
where
\begin{equation} \label{whatisB}
\Lambda = \left[ 1+\left(\frac{r_h}{\rho}\right)^6 \right].
\end{equation}

The conventions used throughout this paper are that capital latin
letters denote generic $D\!=\!11$ indices, world-brane coordinates
are labelled by $x^{\mu}$ where $t=x^0$, and the remaining
coordinates are labelled by $y^m$.  Greek indices are raised and
lowered with $\eta$, lower-case Latin indices are raised and
lowered with $\delta$, and all other metric factors are shown
explicitly.  The radial coordinate 
$\rho\equiv \sqrt{y^p y^p}$; $\eta_{\mu \nu}$ has
signature +1; $\varepsilon^{012}=+1$, etc.  Also, the four-form
field $F$ is obtained from the three-form potential $A$.

The causal structure of the membrane has been discussed in detail
in Ref.~\cite{penrose1}.  It will only be mentioned that the
Penrose diagram is (qualitatively) identical to that for the
extreme Reissner-Nordstr\"{o}m black hole.  That coordinates can be
found that smooth out the singularity at the horizon follows from
the fact that the membrane solution has the structure of
$adS_4{\times}S^7$.  Note that in the isotropic coordinates of
equation (\ref{duffbrane}), the horizon is located at $\rho=0$ and
hence the inside of the membrane cannot be examined in these
coordinates.  However, as is well known, complete spacelike slices
exist that do not penetrate the horizon.


The extremal five-brane solution can be written as~\cite{guven} 
\begin{equation} \label{orig5brane}
\left\{ \begin{array}{lll} 
{{d}}s^2 & = & \Delta^{^{\frac{1}{3}}} \eta_{\mu \nu}
{{d}}x^{\mu} {{d}}x^{\nu} + \Delta^{^{-2}} {{d}}r^2 + r^2
{{d}}\Omega_4^2 \\ F & = & q_m \mbox{\boldmath $\epsilon$}_4
\end{array} \right.
\end{equation}
where $\mbox{\boldmath $\epsilon$}_4$ is the volume element on the
unit $S^4$ surrounding the five-brane, $\Delta$ is given by
\begin{equation}\label{origdelta}
\Delta  =  \left[ 1-\left({\frac{r_h}{r}}\right)^3 \right],
\end{equation}
and $q_m$ and $r_h$ are related by 
\begin{equation}\label{rhqm}
q_m = \pm 9 \, {r_h}^3.
\end{equation}
This can also be put into an isotropic form given by
\begin{equation} \label{iso5brane}
\left\{ \begin{array}{lll} 
{{d}}s^2 & = & \Delta^{^{\frac{1}{3}}} \, \eta_{\mu \nu}
{{d}}x^{\mu} {{d}}x^{\nu} +
\Delta^{^{-\frac{2}{3}}} \, {{d}}y^p {{d}}y^p
\; \\ 
F & = & \frac{q_m}{4!\rho^5} \, \varepsilon_{pqrst}
	\, y^p \, {{d}}y^q
\wedge {{d}}y^r \wedge {{d}}y^s \wedge {{d}}y^t 
\end{array} \right.
\end{equation}
where $r$ and
$\rho \equiv \sqrt{y^p y^p}$
are related by $r^3 =
\rho^3 + {r_h}^3$.  Finally, the field strength tensor, 
$F_{ABCD}$, is
related to the four form, $F$, by
\begin{equation}
F  =  \frac{1}{4} F_{ABCD} {{d}}X^A \wedge 
	{{d}}X^B \wedge 
{{d}}X^C \wedge {{d}}X^D,
\end{equation}
where $X^A = \{x^0,\dots,x^5,y^1,\dots,y^5\}$.

\begin{figure}[hp] 
\epsffile{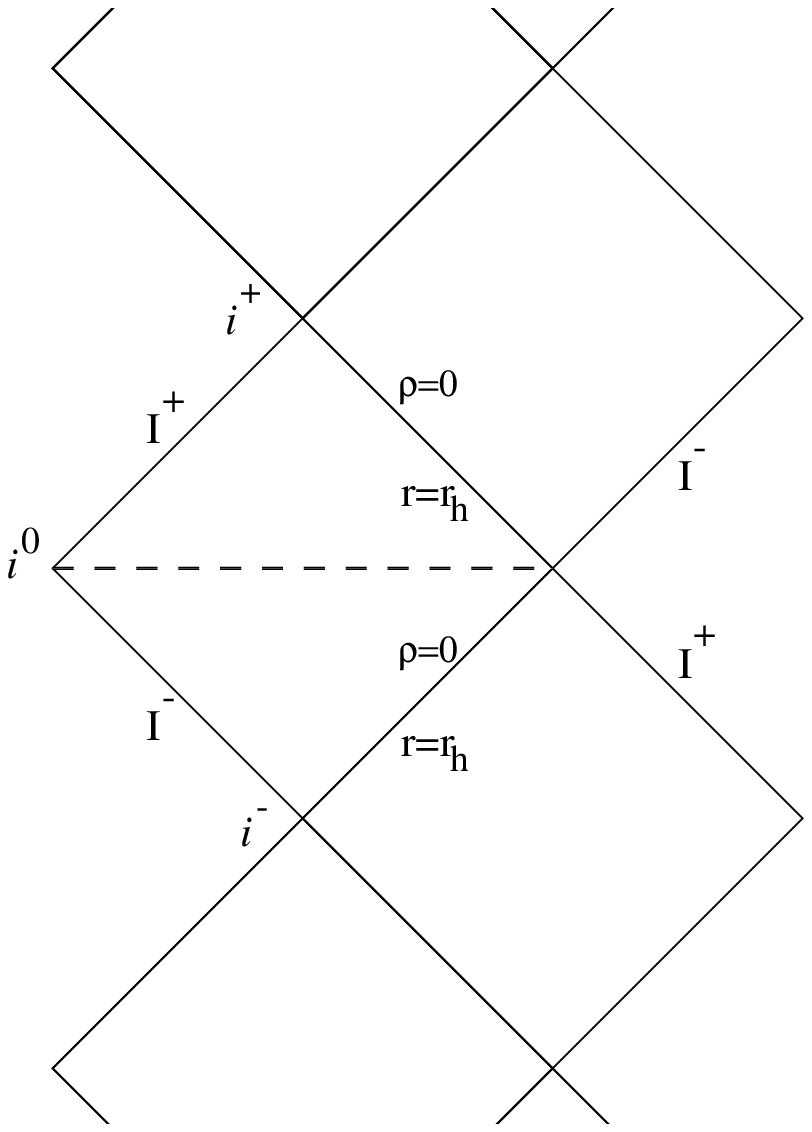}
\caption{Causal Structure of the five-brane.  Each point
represents the product of a four-sphere with a five-plane.  The dashed
line is a spacelike hypersurface.} \label{fig:causalstruct5}
\end{figure}

The causal structure of this solution was investigated
in~\cite{penrose2}.  It was found that the solution can be
continued past the horizon in a totally nonsingular manner.  The
Penrose diagram is given in Fig.~(\ref{fig:causalstruct5}).  It is
clear from this diagram that there are complete spacelike slices
with $r>r_h$ everywhere.  The zero modes of the five-brane will be
defined on such a slice.

\section{Zero Modes and Normalizability} \label{zero}
 
In Ref.~\cite{duff&stelle} it was shown that exactly half of the
$D\!=\!11$ supersymmetries are broken by the membrane solution.
Similar results should hold for the five-brane solution.  Since a
$D=11$ Majorana spinor has 32 real components, it follows that the
supersymmetry breaking leaves 16 (normalizable) fermionic
zero modes, and hence (by supersymmetry), eight (normalizable)
bosonic zero modes.  In the case of the membrane, these will be the
eight translational zero modes.  By spherical symmetry, it is
reasonable to suppose that the effective Lagrangian in static gauge
will simply be (to lowest order)
\begin{equation} \label{guessleff}
{\cal L}_{eff} = -N \eta^{\mu \nu} \partial_{\mu}\lambda^p 
\partial_{\nu}\lambda^p,
\end{equation}
where $N$ is a normalization constant that can be absorbed into
$\lambda^p$, and is therefore meaningless at this order.  It will
be shown that equation~(\ref{guessleff}) holds, with some
arbitrariness in the value of~$N$.

In the case of the five-brane, the situation is more complex.  Here
there are only five translational zero modes.  The remaining three
must be found elsewhere.  It will be shown that these zero modes
exist and are normalizable.  Also, the effective Lagrangian for the
translational zero modes is found to be given by an expression
analogous to equation~(\ref{guessleff}).

\subsection{The membrane solution} \label{membrane}
Obviously, all infinitesimal diffeomorphisms ${y^p}'={y^p}'(y^q)$
generate zero modes; however, only those which correspond to global
translations will be normalizable.  In flat space it is easy to
determine what a global translation is.  This is more difficult in
curved space.  Fortunately, the metric in
equation~(\ref{duffbrane}) is
asymptotically flat; thus global translations must approach
$y^p~\rightarrow~y^p-\epsilon^p$ as $y^p\rightarrow~\infty$, where
$\epsilon$ is a constant vector orthogonal to the membrane+time.
However, there is no well-defined rigid translation away from
$\rho=\infty$.  Specifically, as space becomes curved, the
translation vector $\epsilon$ may be rotated and dilated. In fact,
such diffeomorphisms must be considered in order to cut off the
integral over the off-brane coordinates ($y^m$). 

Thus the Lagrangian is expanded to second order, with
\begin{equation}
\delta g_{AB} = \lambda_{(i)} \mbox{\pounds} _{\epsilon_{(i)}} 
g_{AB}
\end{equation}
and a similar equation for $\delta A_{MNP}$.  Here, $i$ labels the
zero modes (which are eight in number); $\lambda_{(i)}(x^\mu)$ is
the collective coordinate for the i$^{\mbox{th}}$ zero mode; 
the $\epsilon^p_{(i)}(y^m)$ are a set of eight linearly
independent, off-brane vector fields ($\epsilon^\mu=0$) with the
property $\lim_{r \rightarrow \infty}
\epsilon_{(i)}^m=\delta_{mi}$; and \pounds$_{\epsilon}$ denotes
the Lie derivative with respect to the vector field $\epsilon^p$.
Then,
\begin{equation} \label{d2l}
\delta^2 {\cal L}  = \left(\Lambda_{,m} \epsilon^m_{(i)} \partial_n 
\epsilon^n_{(j)}
       + \Lambda\partial_m \epsilon^m_{(i)} \partial_n
\epsilon^n_{(j)} - \frac{1}{2} \Lambda \partial_m \epsilon^n_{(i)}
\partial_m \epsilon^n_{(j)} - \frac{1}{2} \Lambda \partial_m
\epsilon^n_{(i)} \partial_n \epsilon^m_{(j)}\right) \eta^{\mu \nu}
\partial_\mu \lambda_{(i)} \partial_\nu \lambda_{(j)}.
\end{equation}

Integrating equation~(\ref{d2l}) over the off-brane coordinates
gives equation~(\ref{guessleff}), as expected.


\subsection{The five-brane} \label{5braneleff}

Analogously to the membrane, the five-brane will have translational
zero modes.  In this case, there will be five of them.  As eight
(normalizable) bosonic zero modes are expected from supersymmetry, 
we must
find three more zero modes elsewhere.  These extra zero modes are
very similiar to those of the five-brane of type IIA String Theory
investigated in~\cite{worldbranes}.

The equation of motion for small fluctuations of the potential $A$
around a constant background is
\begin{equation}\label{eomsf}
	{{d}}(\hat{*}{{d}}A)+\frac{1}{3}F\wedge {{d}}A
= 0
\end{equation}
where here $F$ is the background field, $A$ is the infinitesimal
fluctuation and $\hat{*}$ is the full 11-dimensional (11D)
Hodge dual.  This is
solved by
\begin{eqnarray}\label{zmsoln}
A & = & \Delta e^{i k_{\mu} x^{\mu}} U \wedge *F \\
  & = & \Delta e^{i k_{\mu} x^{\mu}} U \wedge
\frac{{{d}}r}{r^4}
\end{eqnarray}
where $*$ is the Hodge Dual on the 5D space perpendicular to the
world-brane, $k$ is a constant (anti-)self-dual null vector in two
dimensions tangent to the world-brane and $U$ is a constant
(anti-)self-dual polarization tensor in the four spatial dimensions
orthogonal to $k$ and tangent to the world-brane.  In order to
obtain a normalizable zero mode, $U \wedge k$ must either be
self-dual or anti-self-dual, depending on the sign of $q_m$.  These
zero modes lead to an (anti-)self-dual three-form on the
world-brane, which contains three bosonic degrees of freedom.  As
mentioned above, these zero modes are very similiar to those of the
five-brane of type IIA string theory.

The analysis of the translational zero modes is very similiar to
the analysis of those of the membrane.  Once again it is found that
a suitable choice of $\epsilon^p(y)$ yields normalizable zero modes
whose effective action is
\begin{equation}\label{leff5}
	{\cal L}_{eff} \propto - \eta^{\mu \nu} \lambda^p{}_{,\mu}
\lambda^p{}_{,\nu}.
\end{equation}
in static gauge.


\section{Conclusions} \label{conclusions}
It has been shown that in static gauge, the effective membrane and
five-brane actions are what would be expected from the general theory
of p-branes.  In the case of the membrane, all eight bosonic zero
modes are found to be translational, whereas, for the five-brane, five
are translational and the rest come from an (anti-)self-dual
three-form on the world-brane.  The results for the five-brane are
consistent with a duality between $D=11$ supergravity and $D=10$ type
IIA string theory.  Upon compactification of one of the off-brane
dimensions, one of the translational zero modes of the five-brane
becomes a zero mode of a U(1) field generated by the Kaluza-Klein
mechanism.  As predicted in~\cite{firstfive}, this set of four
translational zero modes, three zero modes due to a self-dual
antisymmetric three-form and one zero mode from a U(1) field, is
exactly the same as the set of zero modes of the five-brane of type
IIA string theory in 10 dimensions investigated in~\cite{worldbranes}.

\section{Acknowledgements}
We thank Andrew Strominger for many useful conversations and
helpful suggestions, and Gary Horowitz for showing us
Ref.~\cite{penrose2}.  We would also like to thank Harald Soleng
for making Ref.~\cite{cartan} available.  One of us (J.M.) thanks
the NSERC and NSF for financial support.
This work was supported in part by DOE Grant No. DOE-91ER40618.

\end{document}